\documentclass[prl,aps,twocolumn,preprintnumbers, showpacs, nofootinbib,superscriptaddress,notitlepage]{revtex4-1}
\usepackage{amssymb,amsthm,amsmath}
\usepackage{textcomp}
\usepackage{color}      
\usepackage{slashed}    
\usepackage{verbatim}
\usepackage[normalem]{ulem} 
\usepackage{bbm,bm}
\usepackage{soul}
\usepackage{rotating}   
\usepackage{multirow}   
\usepackage{simpler-wick}
\usepackage{epsfig}
\usepackage{braket}
\usepackage{hyperref}
\usepackage[splitrule,bottom]{footmisc} 
\usepackage{graphicx}

\usepackage{ragged2e}

\newcommand{\eqal}[1]{\begin{align}#1\end{align}}

\newcommand{\eqsp}[1]{\begin{equation}\begin{split} #1\end{split}\end{equation}}

\begin{document}
 	
\title{
 Impact of parity violation on quantum entanglement and Bell nonlocality
}
\author{Yong Du}\email{yongdu5@sjtu.edu.cn}
\affiliation{Tsung-Dao Lee Institute, Shanghai Jiao Tong University, Shanghai 200240, China} 
\author{Xiao-Gang He}\email{hexg@sjtu.edu.cn}
\affiliation{Tsung-Dao Lee Institute, Shanghai Jiao Tong University, Shanghai 200240, China} 
\affiliation{Key Laboratory for Particle Astrophysics and Cosmology (MOE) \& Shanghai Key Laboratory for Particle Physics and Cosmology, Shanghai Jiao Tong University, Shanghai 200240, China}
\author {Chia-Wei Liu}\email{chiaweiliu@sjtu.edu.cn	}
\affiliation{Tsung-Dao Lee Institute, Shanghai Jiao Tong University, Shanghai 200240, China} 
\author{Jian-Ping Ma}\email{majp@itp.ac.cn}
\affiliation{CAS Key Laboratory of Theoretical Physics, Institute of Theoretical Physics, P.O. Box 2735, Chinese Academy of Sciences, Beijing 100190, China}
\affiliation{School of Physical Sciences, University of Chinese Academy of Sciences, Beijing 100049, China}
\affiliation{School of Physics and Center for High-Energy Physics, Peking University, Beijing 100871, China}
\date{\today}

\begin{abstract}
Quantum entanglement (QE), evidenced by Bell inequality (BI) violations, reveals the nonlocality of nature. Fundamental interactions manifest in various forms, each with distinct effects on QE and BI, but have not yet been studied in depth. We investigate in detail the relationship between QE, Bell nonlocality, and parity-violating interactions in spin-1/2 bipartite systems arising from the decays of spin-0 and spin-1 particles within the quantum field theory~(QFT). Our findings reveal that parity~(P) violation can completely disentangle particle pairs, rendering Bell tests ineffective in distinguishing between classical and quantum theories. In the spin-0 case, complete disentanglement occurs at maximal P violation, which is similarly true for spin-1 decays. Without restrictions from the QFT, the predicted relation between entanglement  and the Bell nonlocality may  no longer be valid and we propose promising methods for testing it. Additionally, we emphasize the previously overlooked influence of magnetic fields within detectors, which alters predictions for QE and Bell nonlocality. This environmental effect induces spurious P  and charge-parity (CP) violations and thus has to be subtracted for genuine P, CP, and Bell tests.
\end{abstract}

\maketitle

{\it Introduction.}---Quantum entanglement (QE) is a fundamental property of quantum mechanics~\cite{Einstein:1935rr} that reveals the inherent nonlocality of nature, famously highlighted by violations of Bell inequalities (BI)\cite{Bell:1964kc}. These inequalities are satisfied by classical local hidden-variable theories but can be violated by quantum systems. Although parity (P) violation is an essential aspect of the Standard Model, current studies on  QE and BI violations  have mostly focused on P-conserving interactions\,\cite{Barger:1988jj, ATLAS:2023fsd, CMS:2024pts, Afik:2020onf,Wu:2024asu}, leaving the role of P violation in these contexts unclear up to now. In addition, many experimental tests of P and charge-parity (CP) violation rely on the presence of QE~\cite{Bernabeu:2012ab, BaBar:2012bwc, BESIII:2021ypr}, yet neither QE nor BI violations have been fully examined in relation to such interactions.

In a broader context, the influence of fundamental particle interactions on the behavior of QE and BI violations has not been extensively investigated. These fundamental interactions manifest in various forms, even beyond the scope of classical and quantum theories, and they play a crucial role in shaping entanglement patterns as well as the degree of BI violations. Conversely, analyzing the characteristics of QE and BI can provide valuable insights into the nature of these underlying interactions.

We demonstrate these important features by studying P-violating interactions on QE and BI for spin-0 and spin-1 decays into two spin-1/2 particle systems in the quantum field theory (QFT). We find that QE and BI crucially depend on the degree of P violation. For P-conserving spin-0 decays, QE and BI violations are maximized, whereas P-violating interactions can completely disentangle bipartite systems, making Bell tests ineffective for distinguishing quantum from classical behavior. This finding holds similarly for spin-1 decays. 
 We find simple ways to parameterize deviations from QFT which can be tested   experimentally.
Moreover, environmental factors like magnetic fields inside detectors can rotate particle spins and mimic P- and CP-violating effects. To accurately access QE and conduct genuine tests of BI, P, and CP violation, it is therefore crucial to account for these ontic environmental influences that were overlooked before in literature.

We start with a general discussion  about QE and BI. The spin-1/2 bipartite system density matrix expanded with  $|\uparrow
   \uparrow  \rangle, | \uparrow   \downarrow\rangle, | \downarrow  \uparrow \rangle, | \downarrow \downarrow  \rangle$
is given by
\begin{eqnarray}\label{density}
	\rho &=&
	\frac{1}{4} \Big ( 
	\boldsymbol{I} _4   +\sum_i {B}_i^{+} \sigma^i \otimes \boldsymbol{I} _2
	\nonumber\\
	&&\qquad	+\sum_j {B}_j^{-} \boldsymbol{I} _2 \otimes \sigma^j  +\sum_{i, j}  {C}_{i j} \sigma^i \otimes \sigma^j \Big )
	\,,
\end{eqnarray}
where $\boldsymbol{I}_n$ is an $n\times n$ unit matrix, and $\sigma^{i,j}$ are the Pauli matrices representing the spin directions. We use boldface symbols to represent square matrices. The size of QE can be quantified by concurrence~\cite{Wootters:1997id}
\begin{equation}
{\cal C} (\rho ) = \max(0, 2 \lambda_{\max} - \text{Tr}( 
{\cal R} 
) )\,,
\end{equation}
where $
{\cal R} = 
\sqrt{ \sqrt{\rho}
(\sigma_y \otimes \sigma_y) \rho^* (\sigma_y \otimes \sigma_y) \sqrt{\rho}  } \,,
$
and $\lambda_{\max}$ is its largest eigenvalue. It satisfies $0\le {\cal C} \le 1$ with ${\cal C}=0$ indicating a vanishing and ${\cal C}=1$ a maximal entanglement, respectively. A relatively stronger requirement of entanglement than the concurrence is the violation of the BI. An example of this is the Clauser-Horne-Shimony-Holt (CHSH) parameter ${\cal B}(\rho)$ which is bounded by 2 in any classical theories. However, in a quantum one, its maximum is $2\sqrt2$ and given by~\cite{Clauser:1969ny,Clauser:1974tg}
\begin{equation}
 	{\cal B}( \rho ) = 2 \sqrt{\mu _1 ^2 + \mu _2 ^2 }\leq 2 \sqrt{2}\,,
\end{equation}
where   $\mu _i^2$  are the eigenvalues of the ${\bf C}^T {\bf C}
$ 
in the order of    $\mu _1 ^2\geq  \mu ^2_2\geq  \mu _3 ^2 $
with $({\bf C})_{ij} = C_{ij} $
and  ${\cal B}(\rho)>2$ signals the Bell nonlocality.

{\it Formalism.}---We set up the formalism for $i\to f_1 \overline{f}_2$ with $i$ either a spin-0 or a spin-1  particle in this section. For a spin-0 particle $h_i$, the amplitude is generically parameterized as
\begin{equation}\label{SP}
M_{\rm scalar}  = 
\bar{f}_1
\left( g_S - g_P \gamma_5 \right)f_2 ,
\end{equation}
with $g_{S\,(P)}$ the (pseudo-)scalar coupling. If $g_S$ and $g_P$ are both nonzero, P is violated. The spinless mother particle would lead to a pure final state given by:
\begin{equation} 
|\Psi \rangle =
\frac{	S+P}{\sqrt{2(|S|^2 + |P|^2)}}|\uparrow \downarrow\rangle
+
\frac{	S-P}{\sqrt{	2(|S|^2 + |P|^2)}}| \downarrow\uparrow
\rangle\,, \label{eq:spin0}
\end{equation}
where the arrows represent the spins of $f_1$ and $\overline{f}_2$ in order along the momentum direction of $f_1$ denoted as $\hat{k}$ throughout this work. Here, $S=\sqrt{m_i^2 - (m_1 + m_2)^2}g_S$, 
$P = \sqrt{m_i^2 - (m_1 - m_2)^2} g_P$, and $m_i$ and $m_{1,2}$ are the masses of $h_i$ and $f_{1,2}$, respectively. It is then straightforward to obtain from eq.~\eqref{eq:spin0} that
\begin{eqnarray}\label{decompose_spin0}
\vec{B}^\pm = \pm \alpha \hat{k}\,,
~~
C_{ij} 
=
\gamma \delta_{ij}- 
( 1 +  \gamma) \hat{k}_i\hat{k}_j
+\beta \epsilon^{ijk} \hat{k}_k\,,
\end{eqnarray}
with the Lee-Yang parameters~\cite{Lee:1957qs}
\eqal{
\alpha = \frac{2 {\rm Re}( S^* P )}{|S|^2 + |P|^2}, \, \beta = \frac{2 {\rm Im}( S^* P )}{|S|^2 + |P|^2}, \, \gamma = \frac{ |S|^2 - |P|^2}{|S|^2 + |P|^2}.
}
\begin{figure}
\includegraphics[width = 0.75\linewidth]{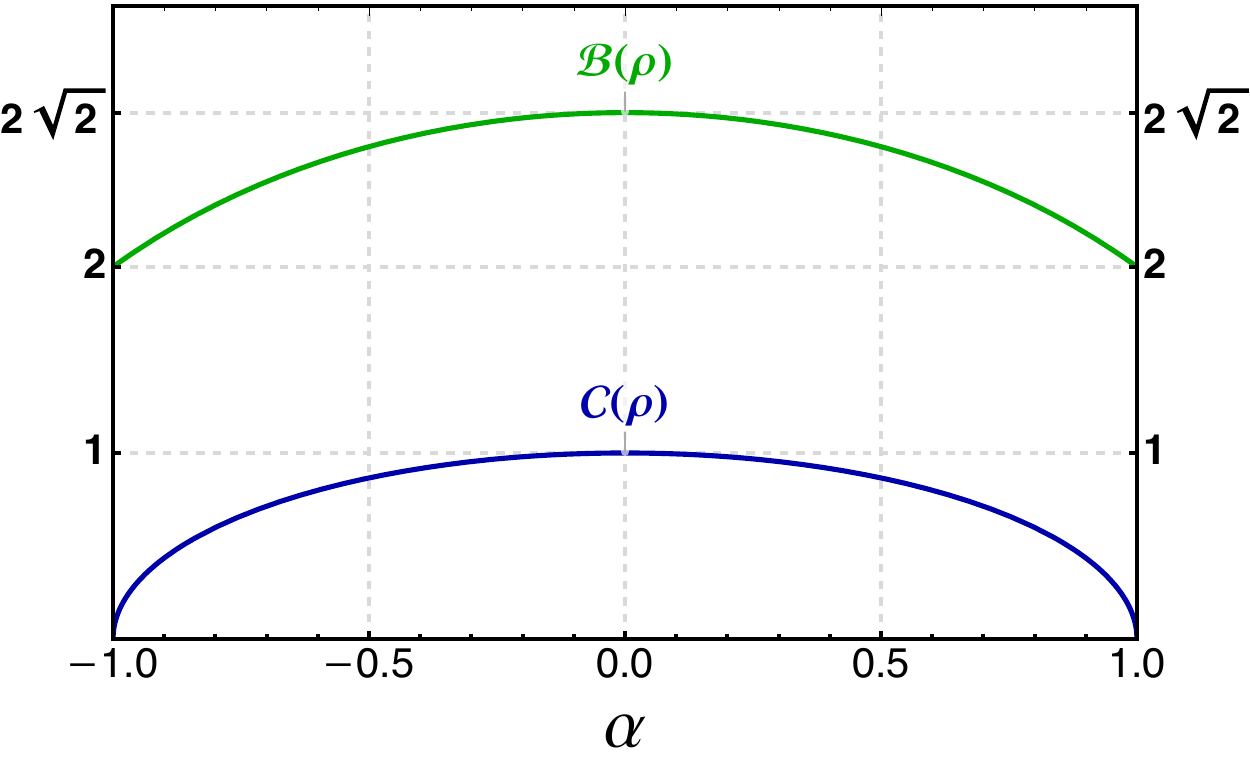}\caption{\raggedright Concurrence ${\cal C}(\rho)$ and the CHSH parameter ${\cal B}(\rho)$ in the spin-0 case. } \label{fig:belle_spin0}
\end{figure}From their definitions, the concurrence and the CHSH parameters in this scenario are given by
\begin{eqnarray}\label{master_spin0}
{\cal C}(\rho )  = \sqrt{
	1 - \alpha^2  
} \,,~~
{\cal B}(\rho)  =  2 \sqrt{ 2 - \alpha^2 }   \,,
\end{eqnarray}
which we also depict in Fig.~\ref{fig:belle_spin0} in blue and green, respectively. As is clearly seen, the daughter particles disentangle completely with ${\cal C}(\rho )=0$ for the maximal value $|\alpha| = 1$, or equivalently $S=\pm P$. At this value ${\cal B}(\rho)=2$ is the boundary of classical and quantum theories and therefore cannot distinguish these two theories. The physical reason for this can be understoodfrom $\ket{\Psi}$ in eq.\,\eqref{eq:spin0}, where only one of the two terms survives, leading to a disentangled bipartite state and suggesting a preference for particular spin combinations and a decrease in QE from P violation. Apart from this special limit, both $\mathcal{C}(\rho)$ and $\mathcal{B}(\rho)$ increase with decreasing $|\alpha|$ and saturate at their respective upper bounds in the optimal limit where $\alpha\to 0$.

For a spin-1 vector $V$, we consider its on-shell production at a lepton collider and its subsequent decay into a spin-1/2 fermion pair denoted as
$e^+e^- \to V\to f \overline{f}$. To address the P-violating effects on QE and BI, we parameterize the decay amplitude as
\begin{eqnarray}
M_{\mathrm{vector}} = \epsilon_\mu \bar u \gamma^\mu( F_V + F_A \gamma_5)v\;,
\end{eqnarray}
where $\epsilon_\mu
$ is the polarization vector of $V$, $u$ and $v$ are the Dirac spinors, $F_{V(A)}$ is the (axial-)vector coupling, and we leave out any possible dipole interactions that can also exist in the most general case. Denoting 3-momenta of $e^-$ and $f$ as $\hat{p}$ and $\hat{k}$, $\vec B^{\pm}$ and $C_{ij}$  are\,\cite{Du:2024jfc}  
\begin{widetext}
\eqal{\label{Cij}
\vec B^{\pm}  = &\, \frac{1}{\bar N}\sqrt{1-y^2_m} \left ( y_m c_\theta \hat p +  (1+ (1-y_m) c_\theta ^2 )\hat k \right ) \text{Re}\left({F_A\over  F_V}\right)\;, \nonumber\\
C_{ij}  = &\, {1\over \bar N} \left [ {1\over 3} \bar N \delta_{ij} + ( 1-(1-y_m^2)\left\vert\frac{F_A}{F_V}\right\vert^2 ) (\hat p_i \hat p_j - {1\over 3}\delta_{ij} ) 
- ( (1-y_m)c_\theta ( 1 - (1+y_m)\left\vert\frac{F_A}{F_V}\right\vert^2 ) 
) (\hat p_i \hat k_j +  \hat k_i \hat p_j - {2\over 3} c_\theta \delta_{ij}) \right.
\nonumber\\
&\, \left.  
+ (1-y_m)\left(1 + c_\theta^2 (1-y_m) 
\right)(\hat k_i \hat k_j - {1\over 3}\delta_{ij})
+ \sqrt{1-y^2_m} s_\theta  \left ( (\hat p_i \hat n_j + \hat n_i \hat p_j) - (1-y_m)c_\theta (\hat k_i \hat n_j + \hat n_i \hat k_j) \right ) \text{Im}\left({F_A \over F_V}\right) \right ]\;, \nonumber\\
\bar N = &\, \frac{1}{2}\left[ 1 + c_\theta^2 + y_m^2 s_\theta^2 
+ (1-y_m^2)(1+c_\theta^2) \left\vert\frac{F_A}{F_V}\right\vert^2   \right],
}
\end{widetext}
where $y_m = 2m_f/m_V$ with $m_{V}$ and $m_f$ masses of the mother and the daughter particles, $\hat{n} = \hat{p}\times \hat{k}/s_\theta $ with $c_\theta\equiv\cos\theta = \hat p \cdot \hat k$ and $s_\theta\equiv\sin\theta$. Here we have neglected P violation in  $e^+e^-\to V$. The full expression with it is given in the end matter of this letter and will be fully taken into account numerically in the next section.  

Interestingly, we note that $C_{ij} \to \hat p_i \hat p_j=\hat{k}_i \hat{k}_j$ with $\theta \to 0$ or $\pi$ such that $C_{ij}$ becomes respecting the parity symmetry. Furthermore, $\mathcal{C}(\rho)\to 0$ and $\mathcal{B}(\rho)\to 2$ in this far forward or backward region and the daughter pair tends to disentangle entirely and the BI also becomes fulfilled. Experimentally, the forward and the backward regions are largely avoided practically for the spin reconstruction of daughter particles due to very large background, we therefore focus our discussion on the phase space away from $\theta\to 0$ and $\pi$ in the following.

\begin{figure}[!t]
  		\includegraphics[width = 0.48 \linewidth]{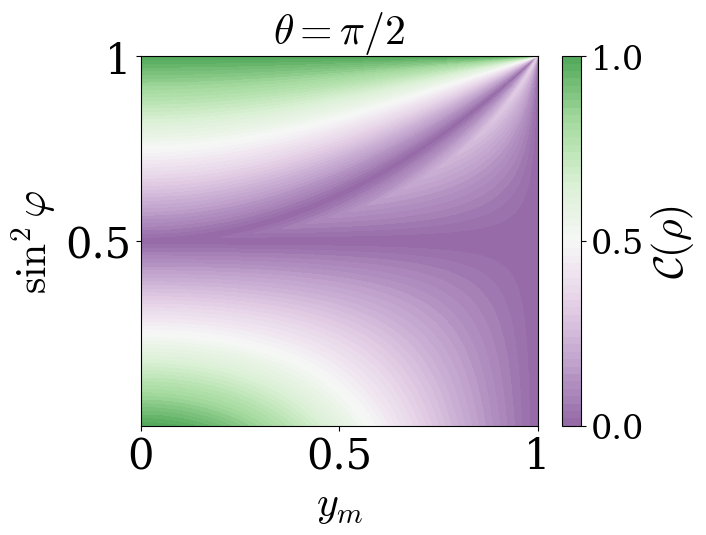}
  		\includegraphics[width = 0.48 \linewidth]{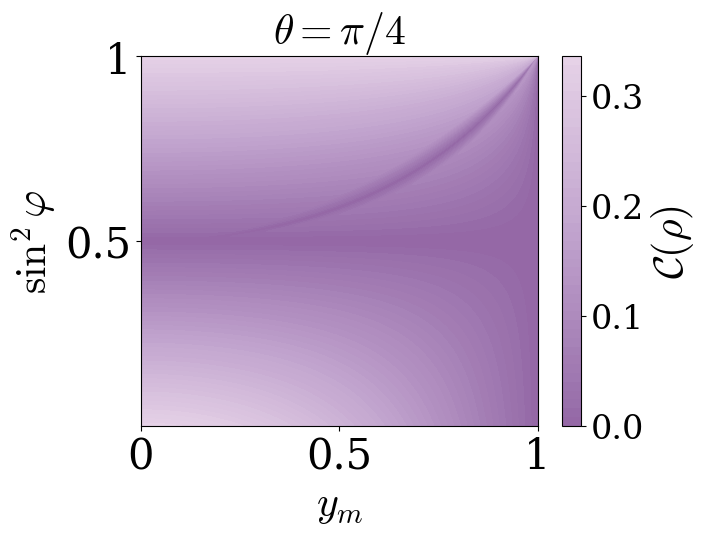}
  		\includegraphics[width = 0.48 \linewidth]{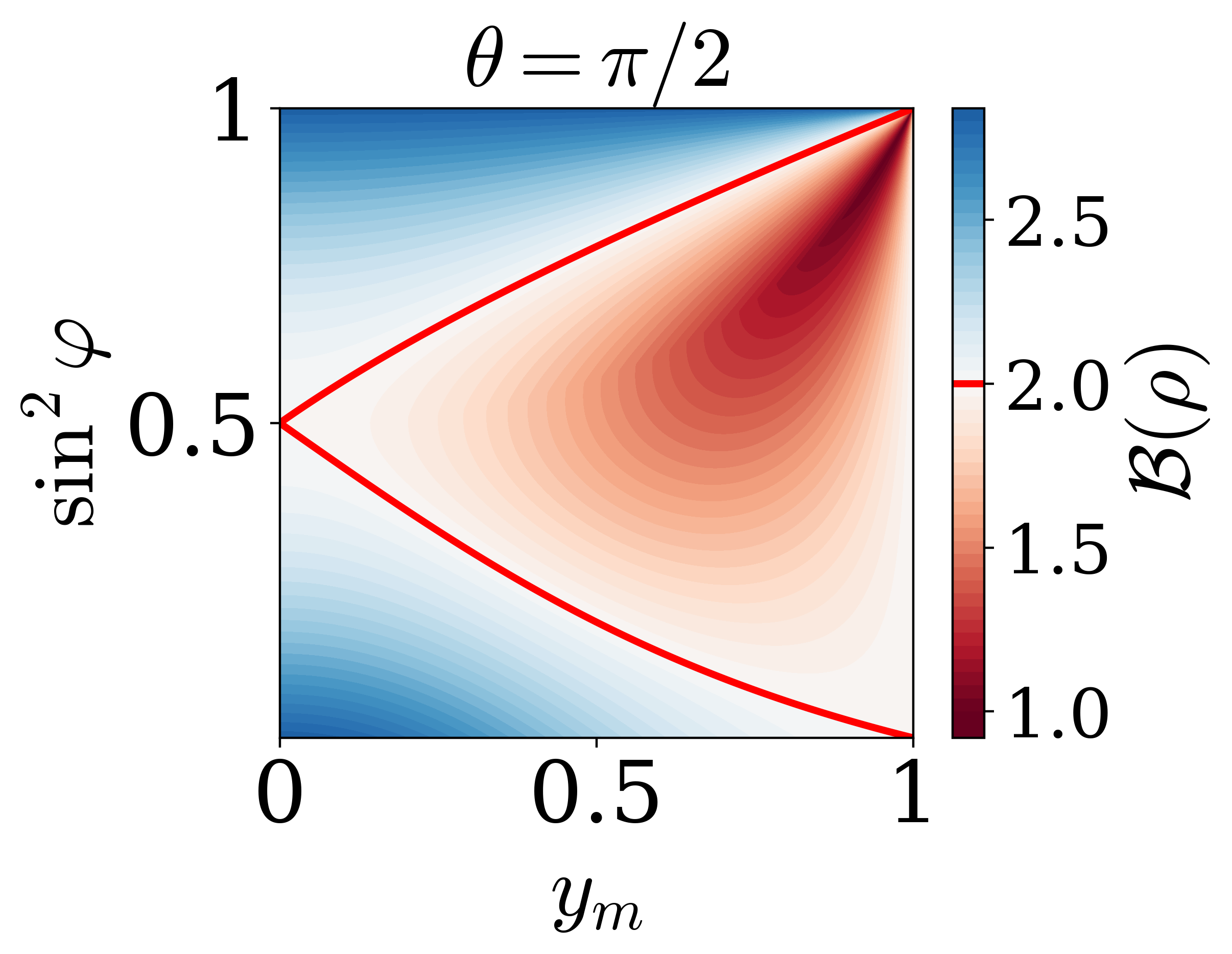}
		\includegraphics[width = 0.48 \linewidth]{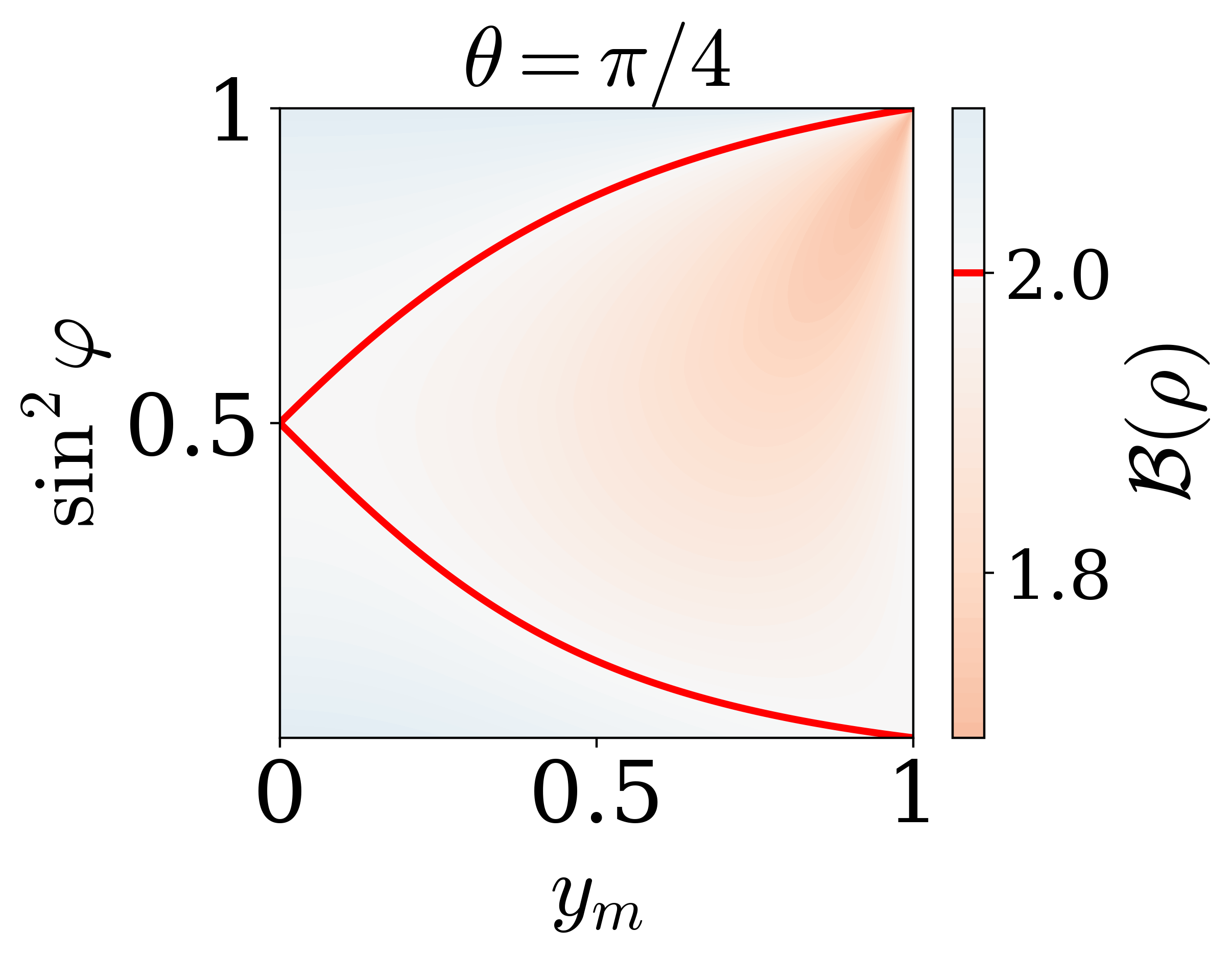}
	\caption{\justifying
Dependencies of ${\cal C}(\rho)$ (upper panels) and ${\cal B}(\rho)$ (lower panels) on $y_m$ and $s_\varphi^2$ at $\theta = \pi/2$ (left panels) and $\theta = \pi/4$ (right panels). The blue regions outside the red line represent the parameter spaces where ${\cal B}(\rho) > 2$,  prohibited in a local hidden-variable theory.
} \label{fig:kappa}
\end{figure}

For simplicity but without sacrificing any physical importance for the discussion below, we assume $F_A/F_V$ to be real from here on and define $\tan\varphi = F_ A / F_V $ along with $( c_\varphi, s_\varphi) = (  \cos \varphi, \sin \varphi)$. The concurrence and the CHSH parameters then become more complicated due to extra freedoms compared with the spin-0 case above. We thus opt to present their numerical results in Fig.\,\ref{fig:kappa}, where the lower row is for $\mathcal{B}(\rho)$ with $\mathcal{B}(\rho)\le2$ as embraced by the two red curves to the right, and the upper row for $\mathcal{C}(\rho)$. Furthermore, similar to $\theta\to0$ or $\pi$ as discussed in the previous paragraph, we observe from Fig.\,\ref{fig:kappa} that the daughter pair also disentangles completely when $y_m\to1$ since $C_{ij} \to \hat p_i \hat p_j$ under this limit. From the above discussions we see that just like the spin-0 cases the level of QE and the size of BI violation are intricately dictated by the interaction forms. In turn, the fundamental interactions can be inferred from the patterns in QE and BI violations. 

On the transverse production plane at $\theta = \pi/2$ with $y_m=0 $, we obtain
\begin{eqnarray}\label{new2}
{\cal C}(\rho)_{\theta= \frac\pi2} 
= \sqrt{
1 - \alpha_f ^2 
},\quad 
{\cal B}(\rho)_{\theta= \frac\pi2}  
= 2 \sqrt{
2 - \alpha_f^2 
}\,.
\end{eqnarray}
Here, $\alpha_f= 2 s_\varphi c_\varphi$ represents the polarization of $V\to f\overline{f}$ and quantifies the size of the P violation. One immediately recognizes its similarity to the spin-0 case in eq.~\eqref{eq:spin0}. Note that  the maximal values of $\mathcal{B}(\rho)$ and $\mathcal{C}(\rho)$ are always achieved on the transverse production plane at $\theta = \pi/2$. With $y_m = 0$, the final state becomes a pure one:
\begin{equation}
|\Psi\rangle_{\theta = \frac{\pi }{2 }}
= \frac{1 }{\sqrt{2}} \left[ 
	(c_\varphi + s_\varphi )  
	\ket{\uparrow\uparrow} 
	+
	(c_\varphi  -  s_\varphi )
	\ket{\downarrow\downarrow} 
\right]\,.\label{eq:spin1}
\end{equation}
For $\tan\varphi = \pm 1$, the final state is disentangled, $\mathcal{B}(\rho)$ and $\mathcal{C}(\rho)$ reach their minima  and fall on the  boundaries of classical or local hidden-variable theories.

In addition, the analogy between eq.\,\eqref{eq:spin0} and eq.\,\eqref{eq:spin1} also renders our discussion on parity in the spin-0 case applicable to this spin-1 scenario but in a different context: both ${\cal C}(\rho)$ and ${\cal B}(\rho)$ will also saturate their upper bounds with $s^2_\varphi\to0$ and $1$ as seen from the upper- and the lower-left corners in the left column of Fig.\,\ref{fig:kappa}. A similar conclusion can be drawn away from the transverse production plane as showcased in the right column of Fig.\,\ref{fig:kappa} with $\theta=\pi/4$. Therefore, it is possible to enhance the effects from parity on entanglement and Bell nonlocality by covering a larger phase space around $\theta=\pi/2$. The average in phase space will in turn change the quantum nature of the bipartite system and thus the tests of BI. We discuss this issue from measurements in the next section.

{\it Measurements of QE and BI}.---The phase space average mentioned in the last section is effectively taken over a set of quantum states represented by $\rho_a$, and the resulting state is a fictitious one\,\cite{Afik:2020onf} which we denote as $\overline{ \rho } = \sum_a^N \rho_a /N$ with $N$ the total number of states. Since $\rho_a$ is obviously frame-dependent, the optimal choice of the frame is therefore the one to maximize the CHSH parameter of the fictitious state
\begin{equation}\label{han}
\overline{{\cal B}}(\rho) = 
\max( {\cal B} (\overline{\rho}' ) ) 
= \frac{2}{N}  \sqrt{
\sum _{i=1,2}
 \left(
\sum_a^N \mu _i^a  
\right)^2 
},
\end{equation}
from an $SU(2)\otimes SU(2)$ rotation $U_a$ such that $\overline{\rho} ' = \sum_a^N  U_a\rho_a U_a^T /N $. $\mu_{1,2}^a$ here are the largest two eigenvalues of ${\bf C}_a$ and can always be taken as positive from an $SU(2)\otimes SU(2)$ rotation. The proof of Eq.~\eqref{han} can be found in Ref.~\cite{Cheng:2024btk}, and we provide an alternative in the end matter.

\begin{figure}[!t]
  		\includegraphics[width = 0.8 \linewidth]{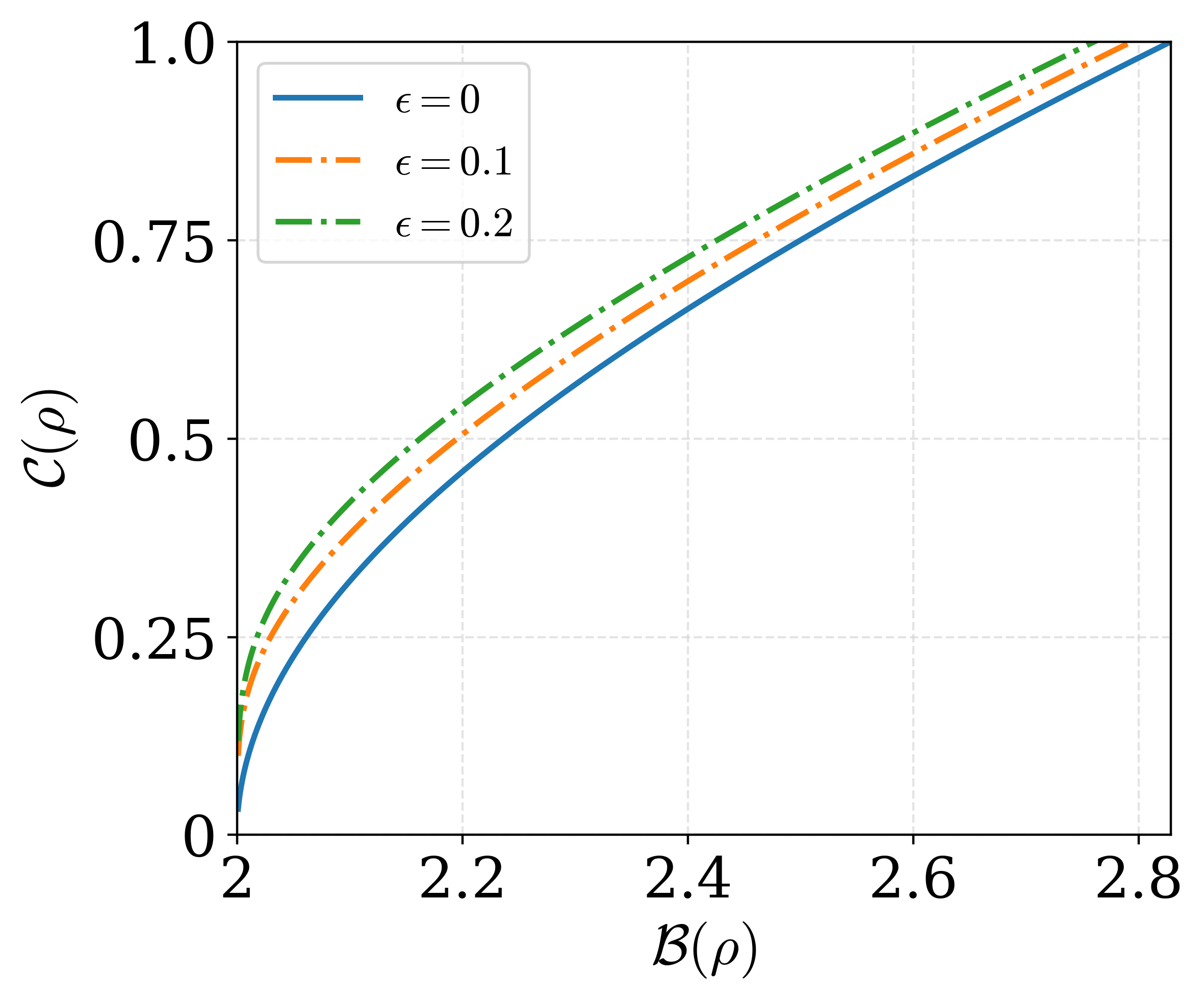}
	\caption{\justifying
Relationship between ${\cal C}(\rho)$ and ${\cal B}(\rho)$. 
The one-to-one correspondence between ${\cal C}(\rho)$ and ${\cal B}(\rho)$ is disrupted by   nonzero $\epsilon $ for illustrations. The blue  line represents the predictions from the QFT, where a nonzero $\epsilon$ cannot be generated under standard QFT assumptions.
} \label{fig:epsi}
\end{figure}

Now we discuss  the most general form of $\rho$ based on fundamental symmetries without   the restrictions of  the QFT for spin-0 and spin-1 systems.
For a spin-0 $h_i$ and its decay $h_i\to f_1\bar f_2$, there is only one  freedom in the rest frame of $h_i$ being the momentum of $f_1$. Therefore, the most general forms for $\vec B^\pm$ and ${\bm C}$ are
\begin{eqnarray}\label{eq17}
	&&\vec{B}^+ = b_{1k} \hat{k}\,,~~~ \vec{B}^- = b_{2k} \hat{k}\,, \nonumber\\
	&&C_{ij} = c_0 \delta_{ij} + c_2 \epsilon_{ijl} \hat{k}_l  + c_5(\hat{k}_i \hat{k}_j - \delta_{ij}/3)\,.
\end{eqnarray}
Due to the absence of any orbital angular momentum along the direction of \(\hat{k}\), the spins of the final states must be opposite, \(\rho(\vec{s}_1 = \vec{s}_2 = \pm \hat{k}) = 0\), where \(\vec{s}_1\) and \(\vec{s}_2\) are the spins of \(f_1\) and \(\overline{f}_2\), respectively. This immediately leads to \(b_{1k} = -b_{2k}\), and \(c_0 = -1 - 2c_5/3\), and the spin-0 system is completely described by three parameters. From eq.~\eqref{eq17}, we have that 
	$\mathcal{B} (\rho)   = 2 \sqrt{2 - b_{1k} ^2 - \epsilon}$ and 
\begin{equation}
\mathcal{C} (\rho)  =
\frac{1}{2}[ 
( \mathcal{B}(\rho)^2 - 4)
(   \mathcal{B}(\rho)^2 - 4 + 4 \epsilon)]^{\frac{1}{4} }\,,
\end{equation}
where \( \epsilon = 1 - b_{1k} ^2 - c_2 ^2 - (1+ c_5)^2 \geq 0 \) to guarantee non-negative probability. The relationship between ${\cal C}(\rho)$ and ${\cal B}(\rho)$ are depicted in Fig.~\ref{fig:epsi}, where the one-to-one correspondence between ${\cal C}(\rho)$ and ${\cal B}(\rho)$ is broken by a nonzero $\epsilon$. In the QFT limit, we have that $c_5 = -1-\gamma$, $  c_2 = \beta $ and $b_{1k} = \alpha$ and \( \epsilon = 0 \) is guaranteed. Therefore, nonzero $\epsilon$   provides a novel method to test the effects beyond the QFT. We emphasize  that theses parameters can be determined by measuring the  cascade decays of $f_1$ and $\overline{f}_2$, described by the end matter of this work. We point out that this examination of $\epsilon \neq 0$ can also be performed using hyperon decays~\cite{ParticleDataGroup:2022pth}.

In this case, we propose testing of the BI with the unexplored weak decay channels $B \to {\bf B}_c \overline{{\bf B}}_c'$, which is well-suited for LHCb and  Belle~II~\cite{LHCb:2014scu,Belle:2018kzz,Belle:2019bgi}. Here, $B = (B^+, B^0, B_s^0)$  and ${\bf B}_c^{(\prime)} = (\Lambda_c^+, \Xi_c^+, \Xi_c^0)$. For all these channels, $\alpha\ne0$ due to parity violation and the violation of BI is expected. We also point it out that it will be interesting to consider neutral $B_s $ mesons that are CP-tagged as $B_{sL}^0/B_{sH}^0$. For these neutral mesons, by ignoring the tiny CP violation from the   mixing, $B_{sL}^0/B_{sH}^0 \to {\bf B}_c \overline{{\bf B}}_c $ conserve CP, leading to $b_{1k}  = c_2 =0 $ in both cases, with $c_5 $ being $-2$ and $0  $, respectively, for $B_{sL}^0$ and $B_{sH}^0$. As a consequence, they are ideal candidates for a maximal violation of the BI.
Similar analysis can be carried out in the $h \to f \overline f $ at the LHC or future colliders. 

On the other hand, for the spin-1 case, $\vec{B}^\pm $ and ${\bf C}$ contain three and nine matrix elements, expanded by $\hat{l}$ with $\hat l = \hat{p }, \hat{k}, \hat n $.  CP symmetry constrains ${\bf C}$ to be symmetric and $\vec{B}^+ = \vec{B}^-$, and one then has nine free parameters in total. In this scenario, we promote $e^+ e ^ - \to Z \to  f \overline{f}$ with $Z$ produced on-shell and consider the decay in the Standard Model  with $f = b,c,\tau$~\cite{Dreiner:1992gt}. The Lagrangian is parameterized as
\begin{equation}\label{Lint}
{\cal L}_{Zff} 
= -\frac{g_L}{2 \cos \theta_W}
Z_\mu
 \bar{f} \gamma^\mu (g_V - g_A \gamma_5) f,
\end{equation}
where $g_L$ is the weak coupling constant, $g_V = I^3_f - 2 Q_f \sin ^2\theta _W$ and $g_A= I^3_f$ with $I^3_f$ and $Q_f$ the isospin and electric charge of $f$. Here, $\theta_W$ is the Weinberg   angle of the weak interaction. According to the heavy quark symmetry, spins of $\Lambda_{b,c}$ contribute exclusively from the heavy quarks $b$ and $c$. Hence, to the leading order, the Lagrangian for $\Lambda_{b,c}$ can be obtained by replacing $f$ by $\Lambda_f$ in eq.~\eqref{Lint}~\cite{Mele:1992kh}, together with an overall hadronization factor $g_\Lambda$ whose explicit number does not enter the discussion for testing QE and the violation of the BI.

\begin{table}[!htb]
\centering
\begin{tabular}{|l|c|ccccccc|}
\hline 
 Processes   &$ -\alpha_f$
&\rule{0pt}{3ex}$\overline{\cal B}_{-1.0} ^{-0.5} $
&$\overline{\cal B}_{-0.5} ^{-0.3} $
&$\overline{\cal B}_{-0.3} ^{-0.1} $
&$\overline{\cal B}_{-0.1} ^{~0.1} $
&$\overline{\cal B}_{~0.1} ^{~0.3} $
&$\overline{\cal B}_{~0.3} ^{~0.5} $
&$\overline{\cal B}_{~0.5} ^{~1.0} $
\\ \hline
$ Z \to \Lambda_b^0  \bar \Lambda_b^0    $         &0.94
&2.01&2.04&2.08&2.10&2.10&2.06&2.01\\ 
$ Z \to \Lambda_c^+  \bar  \Lambda_c^- $    
& 0.70 &
2.03&2.24&2.40&2.49&2.46&2.31&2.05 \\ 
$ Z \to \tau ^- \tau ^+ $      &  0.21
&2.06&2.45&2.69&2.81&2.71&2.47&2.07 \\
\hline
\end{tabular}
\caption{The numerical values of $\overline{{\cal B}}^{\omega_1} _{\omega_2}$ in $Z \to f \overline{f}$, where $\overline{{\cal B}}^{\omega_1} _{\omega_2}$ is the average of $\overline{{\cal B}}$ over the range $\omega_2 > c_\theta  > \omega_1$. $\alpha_{\Lambda_{b,c}}$ are  calculated in the heavy quark limit.}\label{tab:example}
\end{table}

For spin-1 cases, measurements of ${\cal B}$ requires averaging over $\theta$ according to Eq.~\eqref{han}. We define
\begin{eqnarray}
\label{angle_spin-1}
\overline{ {\cal B}}_{  \omega_1}^{\omega_2   
}
	&=& \frac{2}{  	 
		{\cal R}_{\omega_1}^{ \omega_2 }
	}  \sqrt{
		\sum_{i = 1,2} 
		\left[ 
		\int^{ \omega_2 }_{  \omega_1   } 
\mu _i  (c_\theta )   
\left( 		\frac{d\sigma }{ d c_\theta  }\right) 
		d c_\theta 
		\right]^2 
	}\,,
\end{eqnarray}
where $
{\cal R}_{ \omega_1}^{\omega_2  }  =
	\int_{ \omega_1 } ^{\omega_2}    (d \sigma/dc_\theta) dc_\theta\,
$ and $\sigma$ is the scattering rate. Note that $\overline{{\cal B}}_{\omega_1}^{\omega_2} = \overline{{\cal B}}_{- \omega_2}^{- \omega_1}$ up to ${\cal O}(d_J)$, where $d_J$ is the P-violating parameter on the production side, as detailed in the end matter. The numerical values of $\overline {\cal B}$ are documented in Table~\ref{tab:example}. From the table, one observes that a larger \(\alpha_f^2\) leads to  lower  QE, and \({\cal B}(\rho)\) reaches its largest values around \(\theta = \pi/2\), as anticipated. An interesting feature is that the BI is violated in every bin. Genuine P-violating effects are found to be negligible in $J/\psi$ baryonic decays, and $\overline{{\cal B}}_{\omega_1}^{\omega_2} = \overline{{\cal B}}_{- \omega_2}^{- \omega_1}$ holds up to $10^{-4}$, while the magnetic field affects $\overline{\cal B}$ at ${\cal O}(10^{-3})$ as  detailed in the next section. 

{\it External magnetic field effects}.---We now study the effects of the interaction between the environmental external magnetic field and the system on QE and BI. In actual situation, the external magnetic field $\vec{H}$ inside the detector is utilized to reconstruct the momenta of charged particles as well as a measure of their spin orientation. We argue in the following that the existence of a nonzero $\vec{H}$ can introduce a deviation to the density matrix $\rho$. Such effects have been ignored in literature. 

We take the external magnetic field $\vec{H}$ near the production point, which coincides with the beam axis, as $\hat z$ with a magnitude of 1\,tesla\,\cite{BESIII:2009fln,L3:1989aa,ALEPH:1990ndp,OPAL:1990yff,DELPHI:1990cdc}, and the production plane of the charged daughter pair as $\hat{x}-\hat{z}$.  $\vec{H}$ rotates the momentum by 
\eqal{
\hat{l}(t)
= \exp \left(-i \vec{\bm J} \cdot \frac{2 q \vec{H}}{m_V} t \right) 
\hat{l}(0),\label{eq:Rp}
}
with $(\vec{\bm J_i})_{jk}=-i\varepsilon_{ijk}$ the SO(3) generators and  $q$ the electric charge.

For the cases of $\Xi^-$, $\Lambda_c^+$ and $\tau^-$ interested to us, since their Larmor frequencies multiplied by their individual lifetimes are found to be about $1.3\%$, $1.0\times10^{-4}$ and $8.7\times 10^{-4}$, respectively, it then suffices to truncate at the first Magnus expansion for the spin precession. We obtain
\eqsp{
\vec{s}_1(t_1) = e^{-i{\bm \Omega}_1(t_1)} 	\vec{s}_1(0),~~\vec{s}_2(t_2) = e^{i{\bm \Omega}_1(t_2)}	\vec{s}_2(0).
}
with
\eqal{
{\bm \Omega}_1 = \int_0^t dt' \vec{\bm J} \cdot \frac{g q}{2 m_fy_m } \left[  \vec{H}+(y_m-1 ) \vec{H} \cdot \hat{k} \hat{k}(t') \right],
}
where $g$ is the gyromagnetic ratio, and note that \(\vec{s}_1\) and \(\vec{s}_2\) rotate oppositely due to their opposite magnetic dipole moments.

The modified $\rho$ is then obtained upon replacing momenta and spins at the production time by those at the decay time, and its time average is calculated with a Gaussian distribution $p(t_1,t_2)$ given by
\eqal{
	p(t_1,t_2) = &\, \frac{1}{2\pi \sigma_{\rm TOF}^2} e^{-\frac{(t_1 - \tau)^2+(t_2 - \tau)^2}{2\sigma_{\rm TOF}^2}},
}
with $\sigma_{\rm TOF}$ the time resolution of the time-of-flight (TOF) system of the detector which  is $65/2$ ps for BESIII\,\cite{Guo:2017sjt} and 300/2\,ps as the typical value for LEP detectors\,\cite{L3:1989aa,ALEPH:1990ndp,OPAL:1990yff,DELPHI:1990cdc}, such that
\eqal{
	\bar \rho = &\, \int dt_1 dt_2\, \rho(t_1, t_2) p(t_1,t_2),\,
	\bar{C}_{ij} = {\rm Tr}\left[ \bar \rho \sigma_i\otimes\sigma_j\right].
}

\begin{figure}
\begin{center}
\includegraphics[width = 0.75 \linewidth]{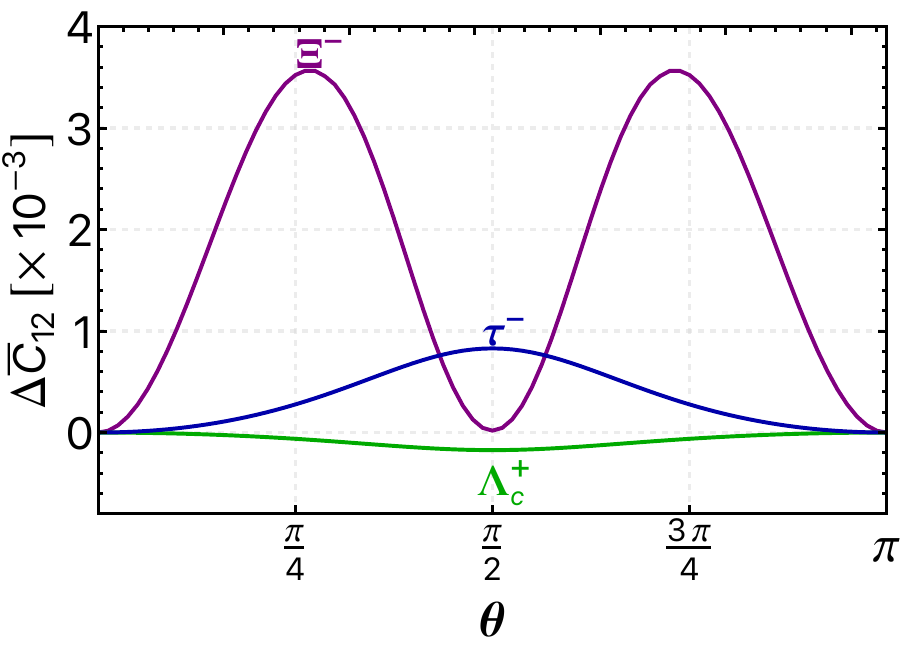}\\
\caption{\justifying
Spurious P and/or CP violating effects from the external magnetic field for $J/\psi\to\Xi^-\bar\Xi^+$ (purple), and $Z\to\Lambda_c^+\bar\Lambda_c^-$ (green) and $\tau^-\tau^+$ (blue), where $\theta$ is the production angle of the final state particle.}\label{fig:C12}
\end{center}
\end{figure}

Numerically, we find the influence on $\mathcal{B}$ from the external $\vec{H}$ field marginal. Modifications to the spin correlation matrix $C_{ij}$ are found important. We illustrate this point in Fig.\,\ref{fig:C12}, where the $y$-axis is defined as
\eqal{
\Delta \bar C_{12}\equiv \bar C_{12} - \bar C_{21}.
}
Since $C_{12}=C_{21}$ in the P- and CP-conserving limit, $\Delta \bar C_{12}$  gives a net measure on the magnetic effect. 
As is clearly seen from the plot, the presence of a non-vanishing magnetic field can induce spurious P and/or CP violating effects, which in turn produces a nonzero $\Delta \bar C_{12}$ of $\mathcal{O}(10^{-3})$ for $J/\psi\to\Xi^-\bar\Xi^+$, and of $\mathcal{O}(10^{-4})$ for $Z\to\Lambda_c^+\bar\Lambda_c^-,\,\tau^-\tau^+$ as shown in Fig.\,\ref{fig:C12}. On the other hand, since $C_{ij}$ is directly related to the differential angular distribution as shown in the end matter, such a spurious effect has to  be isolated from the fitting to obtain a genuine determination of P and/or CP violation.  Moreover, if this effect is not subtracted in the analysis and the bipartite system is assumed to be isolated as before, one may falsely draw the conclusion of invalidity of the QFT as similarly depicted in Fig.~\ref{fig:epsi}.

The nonzero $\Delta \bar C_{ij}$ are detectable through the cascade decays of the daughter fermions, \( f \to f' X \) and \( \overline{f} \to \overline{f}' X \) with \( f' \)  a fermion and \( X \)  the rest of the particles. Ideal choices of the cascade decays, for instance, are \( \Xi^- \to \Lambda \pi^-, ~ \Lambda_b^0 \to \Lambda_c^+ \pi^-, ~ \Lambda_c^+ \to \Lambda \pi^- \), and \( \tau^- \to \pi^- \nu_\tau \)~\cite{LHCb:2024tnq}, as described in the end matter. Measurements would be able to  carry out at next-generation colliders such as STCF\,\cite{Charm-TauFactory:2013cnj,Achasov:2023gey}, CEPC\,\cite{CEPCStudyGroup:2018ghi} and FCC-ee\,\cite{FCC:2018evy}.   

{\it Acknowledgements}---This work was partially supported by the Fundamental Research Funds for the Central Universities, by NSFC grant numbers  
11821505, 11935017, 12075299, 12090064, 
12205063, 12347116 and 12375088, by the Strategic Priority Research Program of Chinese Academy of Sciences  grant   number   XDB34000000, by CPSF grant numbers 2023M732255, 2023M742293 and  GZC20231613.

\bibliographystyle{utcaps_mod}
\bibliography{ref} 

\clearpage 

\onecolumngrid

\begin{center}
 	\textbf{{\normalsize End matter  for “Impact of parity violation on quantum entanglement and Bell nonlocality"}} 
 \end{center}

\appendix
\section*{
Maximum of ${\cal B}(\overline{\rho}) $ for a fictitious state 
}
\noindent \underline{{\bf Lemma:}}
There exist matrices \({\bf R}_\pm\) such that \({\bf R}_+ \mathbf{C} {\bf R}_-=\)diag$(\mu_1,\mu_2\cdots \mu_N)$ with \(\mathbf{C}\) an arbitrary \(N \times N\) real matrix  and \({\bf R}_\pm \in SO(N)\). Furthermore,   \(\mu _{i} \geq |\mu _{i+1}|\) for \(i = \{1, 2, \dots, N-1\}\).

Since \(\mathbf{C}^T \mathbf{C}\) is semipositive and symmetric, there exists a matrix \({\bf R} \in SO(N)\) such that \({\bf R}^T \mathbf{C}^T \mathbf{C}{\bf R}=\)diag$(\mu_1^2,\mu_2^2\cdots \mu_N^2)$ with the hierarchical order \(\mu _1^2 \geq \mu _2^2 \geq \cdots \geq \mu _N^2\). Given that \(({\bf R} \mathbf{C} {\bf R})^T ({\bf R} \mathbf{C} {\bf R})\) is diagonal, we must have
\begin{equation}\label{27}
 {\bf R} \mathbf{C} {\bf R}  = 
\left(
\begin{array}{c}
	\mu _1 \hat{v}_1^T\,,~
	\mu _2 \hat{v}_2^T\,,\cdots
,~	\mu _N \hat{v}_N^T
\end{array}
\right)\,,
\end{equation}
where \(\mu _i\) are real constants,
$\hat{v}_i
= ((\hat{v}_i)_1,
~(\hat{v}_i)_2, \cdots
,~(\hat{v}_i)_N) 
$ and 
\(\hat{v}_i \cdot \hat{v}_j = \delta_{ij}\). Define
\begin{equation}
{\bf R}^{\prime T}  = 
\left(
\begin{array}{c}
	\text{sgn}(\mu _1) \hat{v}_1 ,\,
	\text{sgn}(\mu _2) \hat{v}_2
	, \cdots
,~ 
	\left(\prod_{i=1}^{N-1} \text{sgn}(\mu _i)\right) \hat{v}_N
\end{array} 
\right)
\in SO(N)
\,.
\end{equation}
With $\hat{v}_i\cdot \hat{v}_j = \delta_{ij}$,
we see 
  that \({\bf R}' ({\bf R} \mathbf{C} {\bf R}) = \text{diag}(|\mu _1|, |\mu _2|, \cdots, \left(\prod_{i=1}^{N-1} \text{sgn}(\mu _i)\right) \mu _N)\). By identifying \({\bf R}' {\bf R} = {\bf R}_+\) and \({\bf R} = {\bf R}_-\), we prove the lemma.
  The   proof also provides a concrete method to obtain ${\bf R}_\pm$ from    ${\bf C}$.
\hfill $\blacksquare$

The spin-correlation matrix of a fictitious density matrix  is given by 
\begin{equation}
	\overline{{\bf C}}  =  \frac{1}{N} \sum _a ^N  {\bf C} _a 
	=  \frac{1}{N} \sum _a ^N  {\bf R}_+^a    {\bf C}_{\text{diag}} ^a   {\bf R}_-^a
	\,,
\end{equation}
where $  {\bf C} ^a_ {\text{diag}} = $diag$(\mu _1^a, \mu _2^a,\mu  _3^a ).$
We have used the above lemma to diagonalize ${\bf C}_a$ and chosen 
$
\mu  _1 ^a \geq  \mu _2 ^a \geq  |\mu _3^a| $.
Likewise, we 
 diagonalize $\overline {\bf C} = $diag$(\mu _1,\mu _2 ,\mu _3)$  
with 
$\mu _1 \geq \mu _2 \geq |\mu _3|$. 
It is important to note that $\mu_{1,2,3}$ depend on the 
chosen basis or equivalently the choice of ${\bf R}_\pm ^a$.
In the following 
$\mu _{1,2,3}$  are named as $\bar \mu_{1,2,3}$ when   ${\bf R}_\pm ^a = {\bf I}_3$ for all $a$. Eq.~\eqref{han} 
is equivalent to the statement of that 
$\overline{\mu}_1^2 + 
\overline{\mu}_2^2\geq \mu_1 ^2 + \mu_2 ^2 
$, of which we aim to prove now.

Since $\mu ^a_1\geq \mu ^a_2 \geq |\mu  _3^a |$,  we have that 
\begin{eqnarray}
({\bf R}_+^a  
 {\bf C}_{\text{diag}} ^a 
{\bf R}_ - ^a  )_{11}  \leq 
 \mu  _1 ^a   
	({\bf R}_+^a  {\bf R}_ - ^a  )_{1 1}   
\,. 
\end{eqnarray}
Summing the index of $a$ and note $1 \geq 
({\bf R}_+^a  {\bf R}_ - ^a  )_{1 1}   
$, we arrive at $\mu_1\leq \overline{\mu}_1.$
For  $\mu _1 + \mu _2$, we find 
	\begin{eqnarray}\label{7}
		&&	
({\bf R}_+^a  
 {\bf C}_{\text{diag}} ^a 
{\bf R}_ - ^a  )_{11} 
+
({\bf R}_+^a  
 {\bf C}_{\text{diag}} ^a 
{\bf R}_ - ^a  )_{22} 
		\leq   
		( 	\mu  _ 1 ^a - \mu  _2 ^a ) 
		\left( 
		({\bf R}_+^a)  _{11 } 
		( {\bf R}_ - ^a  )_{11}  
		+ 
		({\bf R}_+^a)  _{21 } 
		( {\bf R}_ - ^a  )_{12}  \right)  \nonumber \\
		&+& 
		\mu  _ 2 ^a \left\{ 
		\left[
		\left| ({\bf R}_+^a)  _{13} 
		( {\bf R}_ - ^a  )_{31} \right|   
		+ 
		\sum _{k=1}^2 
		({\bf R}_+^a)  _{1k } 
		({\bf R}_ - ^a  )_{k1}  
		\right]
		+ 
		\left[ 
		\left| ({\bf R}_+^a)  _{23} 
		( {\bf R}_ - ^a  )_{32} \right| +
		\sum _{k=1}^2 
		({\bf R}_+^a)  _{2k } 
		( {\bf R}_ - ^a  )_{k 2}  \right] \right\} \,.
	\end{eqnarray}
The above inequality may seem tedious, but it can be straightforwardly proved by expanding all the summations and using the relation:
$\mu_2^a | ({\bf R}_-^a)_{i3} ({\bf R}_+^a)_{3i} | \geq |\mu_3^a| ({\bf R}_-^a)_{i3} ({\bf R}_+^a)_{3i}$
for \(i = 1, 2\). The two square brackets in the second line of eq.~\eqref{7} can be identified as \( ({\bf R}_+^a {\bf R}_-^a)_{ii} \) if
$( {\bf R}_-^a )_{i3} ({\bf R}_+^a)_{3i} \geq 0$,
or \( ({\bf R}_+^a \text{diag}(1,1,-1) {\bf R}_-^a)_{ii} \) if
$( {\bf R}_-^a )_{i3} ({\bf R}_+^a)_{3i} < 0.$
In either case, we can substitute the second line of eq.~\eqref{7} with \( 2 \mu_2^a \), since
$ 
1 \geq ({\bf R}_+^a {\bf R}_-^a)_{ii} 
\quad \text{and} \quad 
1 \geq ({\bf R}_+^a \text{diag}(1,1,-1) {\bf R}_-^a)_{ii}
$
due to the fact that the matrix elements of the \(O(N)\) rotation group cannot be larger than 1. By the same reasoning, we have
$ 
1 \geq \left( ({\bf R}_+^a)_{11} ( {\bf R}_-^a )_{11} + ({\bf R}_+^a)_{21} ( {\bf R}_-^a )_{12} \right),$
and replace the right-hand side of eq.~\eqref{7} by \( \mu_1^a + \mu_2^a \). Summing over the index \(a\), we find \( \bar{\mu}_1 + \bar{\mu}_2 \geq \mu_1 + \mu_2 \). Together with \( \bar{\mu}_1 \geq \mu_1 \), we arrive at the desired inequality:
$
\bar{\mu}_1^2 + \bar{\mu}_2^2 \geq \mu_1^2 + \mu_2^2$.

\vspace{-0.2cm}
\section*{
Full expression of density matrix for spin-1 decays 
}
\vspace{-0.2cm}
The P-violating effects in \( e^+ e^- \to V \) production, which influence the polarization sum of $V$, can be parameterized by a term proportional to $d_J$:
$
\tilde \rho_{ij} = \ \delta_{ij}/3 - i d_J \epsilon_{ijk} \hat{p}^k - \left( \hat{p}^i \hat{p}^j -  \delta^{ij}/3 \right)/2.$
When $d_J$ is nonzero, the corresponding density matrix is given by:
\eqal{\label{Cij}
\vec B^{\pm}  = &\,
\frac{1}{\bar N}\sqrt{1-y^2_m} \left ( y_m c_\theta \hat p +  (1+ (1-y_m) c_\theta ^2 )\hat k \right ) \text{Re}\left({F_A\over  F_V}\right) \nonumber\\
& +
 \frac{d_J}{\bar N} \left( 
 2y_m   \hat p +  2 (1-y_m)   c_\theta \hat k + 2 y_m d_J \sqrt{1-y_m^2} s_\theta \text{Im}\left({F_A \over F_V}\right) \hat{n}  + 2  c_\theta (1-y_m^2) \left\vert\frac{F_A}{F_V}\right\vert^2 \hat{k}\right)\,,  \nonumber\\
C_{ij}  = &\, {1\over \bar N} \left [ {1\over 3} \bar N \delta_{ij} + ( 1-(1-y_m^2)\left\vert\frac{F_A}{F_V}\right\vert^2 ) (\hat p_i \hat p_j - {1\over 3}\delta_{ij} ) 
- ( (1-y_m)c_\theta ( 1 - (1+y_m)\left\vert\frac{F_A}{F_V}\right\vert^2 ) 
) (\hat p_i \hat k_j +  \hat k_i \hat p_j - {2\over 3} c_\theta \delta_{ij}) \right.
\nonumber\\
&\, \left.  
+ (1-y_m)\left(1 + c_\theta^2 (1-y_m) 
\right)(\hat k_i \hat k_j - {1\over 3}\delta_{ij})
+ \sqrt{1-y^2_m} s_\theta  \left ( (\hat p_i \hat n_j + \hat n_i \hat p_j) - (1-y_m)c_\theta (\hat k_i \hat n_j + \hat n_i \hat k_j) \right ) \text{Im}\left({F_A \over F_V}\right) \right ]\;, \nonumber\\
 & 
 + 
 \frac{d_J}{\bar N}
  \text{Re}
 \left({F_A\over  F_V}\right)  \left[  
 2 y_m   \sqrt{ 1 - y_m^2 } \text{Re}
 \left(\hat p_i \hat k_j +  \hat k_i \hat p_j - {2\over 3} c_\theta \delta_{ij}\right)  
+   4   (1-y_m)\sqrt{1-y_m^2}c_\theta
 \left(\hat k_i \hat k_j - {1\over 3}\delta_{ij}\right) \right]  
  \nonumber\\
\bar N = &\, 
 \frac{1}{2}\left[ 1 + c_\theta^2 + y_m^2 s_\theta^2 
+ (1-y_m^2)(1+c_\theta^2) \left\vert\frac{F_A}{F_V}\right\vert^2   \right]  +
4 d_J \sqrt{1-y_m^2} c_\theta \text{Re}\left({F_A\over  F_V}\right) \,.}
For an on-shell $Z$, both the P-conserving and P-violating interactions are dominated by the $e^+e^-$ interaction with the $Z$ boson, and $d_J =  -(1 - 4 \sin^2\theta_W) /(2 - 8 \sin^2\theta_W + 16 \sin^4\theta_W) $. For $V = J/\psi$, the P-conserving interaction arises from photon exchange at $\sqrt{s} = m_{J/\psi}$, while the P-violating interaction is again from $Z$ exchange, which gives $d_J = \sqrt{2} m^2_{J/\psi}G_F(3 - 8\sin^2\theta_W)/(32\pi \alpha_{EM})$.

\vspace{-0.2cm}
\section*{On extracting $C_{ij}$ in experiments} 
\vspace{-0.2cm}

Consider the cascade decays of $i \to f(\to f' X) \overline{f}(\to \overline{f}' X),$ where $f'$ is a spin-1/2 fermion and $X$ represents the rest of the particles. The 3-momenta of $f'$ and $\overline{f}'$ are denoted as $\hat{a}$ and $\hat{b}$ in the rest frames of $f$ and $\overline{f}$, respectively. The differential distributions are related to the density matrix as  
\begin{equation}
\frac{\partial }{\partial c_\theta} \left( 
\frac{
\partial^4  N
}{
\partial \phi^a _i
\partial \hat{a}_i
\partial \phi^b _j
\partial \hat{b}_j
} \right) 
= 
\frac{1}{16\pi^2 } \frac{\partial N}{\partial c_\theta} 
\left( 
1 +
\alpha ' \vec{B}^+ \cdot \hat{a}
+ 
\overline{\alpha } ' \vec{B}^- \cdot \hat{b}
+ 
\alpha ' \overline{\alpha }' \left( 
\hat{a} \cdot {\bf C} \cdot \hat{b} \right) \right)  \,,
\end{equation}
where $i, j = x,y ,z$, $N$ stands for the number of observed events, and $\alpha'~(\overline{\alpha}' )$ is the polarization fraction of $f'~(\overline{f}')$ in $f \to f' X$ $(\overline{f}\to \overline{f} 'X)$. Here, $\phi^a_i$ and $\phi_j^b$ represent the azimuthal angles of $\hat{a}$ and $\hat{b}$ about $i$ and $j$, respectively. 

Explicitly, for $i = z$, we have  
\begin{equation}\label{33}
(\hat{a}_x,\hat{a}_y, \hat{a}_z ) 
= (\cos\phi_z^a \sin \theta_z^a  ,
\sin \phi_z^a \sin \theta_z^a,  \cos \theta_z^a )      
\end{equation}
and 
$d\phi_a^z  d \cos\theta_z^a$ corresponds to the differential surface area expanded on a unit sphere. 
On the other hand, if we choose $i = x$ or $i = y$ instead, we replace $(\hat{a}_x,\hat{a}_y,\hat{a}_z)$ in eq.~\eqref{33} by $(\hat{a}_y,\hat{a}_z,\hat{a}_x)$ or $(\hat{a}_z,\hat{a}_x,\hat{a}_y)$, while replacing $(\theta_z^a, \phi_z^a)$ with $(\theta_x^a, \phi_x^a)$ or $(\theta_y^a, \phi_y^a)$. The same arguments apply to $\hat{b}$. The choice of $i$ and $j$ does not affect the expression on the right-hand side of eq.~\eqref{33} due to the unity of the Jacobian. 
To extract $C_{ij}$, we integrate over $\phi_i^a$ and $\phi_j^b$ and find  
\begin{equation}\label{Diff}
    \frac{\partial }{\partial c_\theta} \left( 
\frac{
\partial^2  N
}{
\partial \hat{a}_i
\partial \hat{b}_j
} \right) 
= 
\frac{1}{4} \frac{\partial N}{\partial c_\theta} 
\left( 
1 +
\alpha '  
 B^+_i \cos \theta_i^a
+ 
\overline{\alpha } ' 
  B^-_i \cos \theta_j^b
+
\alpha ' 
\overline{\alpha } '
\cos \theta_i^a \cos \theta_j^b 
C_{ij} 
\right)  \,.
\end{equation}
We do not sum over $i$ and $j$. The integration over a bin of $c_\theta$, $\hat{a}_i$, and $\hat{b}_j$ on the left-hand side of Eq.~\eqref{Diff} can be obtained from experiments. 
There are many ways to extract $B_i^\pm$ and $C_{ij}$. We provide the simplest two-bin scenario:  
\begin{equation}
\overline{C}_{ij } = 
 \frac{4}{N_{\omega_1}^{\omega_2}} \int_{\omega_1}^{\omega_2} 
dc_\theta
\left( 
\int_0^1 d\cos \theta_i^a - \int_{-1}^0 d\cos \theta_i^a \right) 
\left( 
\int_0^1 d\cos \theta_j^b - \int_{-1}^0 d\cos \theta_j^b \right) 
\frac{\partial}{\partial c_\theta} \left( 
\frac{
\partial^2 N
}{
\partial \hat{a}_i
\partial \hat{b}_j
} \right)\,.
\end{equation}
Here, $N_{\omega_1}^{\omega_2}$ is the total number of observed events in the range $\omega_2 \geq c_\theta \geq \omega_1$, and $\overline{C}_{ij}$ is the average spin-correlation matrix element over the same range. For spin-0 decays, $C_{ij}$ does not depend on $c_\theta $, and we have $\overline{C}_{ij} = C_{ij}.$ 
For 
$\Lambda_b^0 \to \Lambda_c^+ \pi^-$, 
$\Lambda_c^+ \to \Lambda \pi ^+$ and 
$\tau^- \to  \nu_\tau \pi^-, $
CP is conserved in the cascade decays and we have 
$\alpha '= -\overline{\alpha} '$. Numerically, they are found to be 
  $\alpha' = -1.003 \pm 0.008 \pm 0.005$,
$\alpha'  = -0.785\pm 0.007$
and $\alpha '=-1 $, respectively~\cite{LHCb:2024tnq}.

\end{document}